\documentstyle[psfig,12pt]{article}
\def\TL{\hfil$\displaystyle{##}$}
\def\TR{$\displaystyle{{}##}$\hfil}

\def\seqalign#1#2{\vcenter{\openup1\jot
  \halign{\strut #1\cr #2 \cr}}}

\def\comment#1{}
\def\fixit#1{}

\def\tf#1#2{{\textstyle{#1 \over #2}}}

\def\mop#1{\mathop{\rm #1}\nolimits}

\def\coth{\mop{coth}}
\def\csch{\mop{csch}}

\def\tr{\mop{tr}}

\def\lsim{\mathrel{\mathstrut\smash{\ooalign{\raise2.5pt\hbox{$<$}\cr\lower2.5pt\hbox{$\sim$}}}}}
\def\gsim{\mathrel{\mathstrut\smash{\ooalign{\raise2.5pt\hbox{$>$}\cr\lower2.5pt\hbox{$\sim$}}}}}

\def\sqr#1#2{{\vcenter{\vbox{\hrule height.#2pt
         \hbox{\vrule width.#2pt height#1pt \kern#1pt
            \vrule width.#2pt}
         \hrule height.#2pt}}}}

\def\href#1#2{#2}  

\def\lbldef#1#2{\expandafter\gdef\csname #1\endcsname {#2}}
\def\eqn#1#2{\lbldef{#1}{(\ref{#1})}%
\begin{equation} #2 \label{#1} \end{equation}}
\def\eqalign#1{\vcenter{\openup1\jot
    \halign{\strut\span\TL & \span\TR\cr #1 \cr
   }}}

\def\Li{\mop{Li}}
\begin{document}
\baselineskip=15.5pt
\pagestyle{plain}
\setcounter{page}{1}
\renewcommand{\thefootnote}{\fnsymbol{footnote}}
\begin{titlepage}

\begin{flushright}
HUTP-{\sl A}072\\
hep-th/9810225
\end{flushright}
\vfil

\begin{center}
{\huge Thermodynamics of spinning D3-branes}
\end{center}

\vfil
\begin{center}
{\large Steven S. Gubser}\\
\vspace{1mm}
Lyman Laboratory of Physics \\
Cambridge, MA  02138 \\
\end{center}

\vfil

\begin{center}
{\large Abstract}
\end{center}

\noindent
 Spinning black three-branes in type IIB supergravity are
thermodynamically stable up to a critical value of the angular
momentum density.  Inside the region of thermodynamic stability, the
free energy from supergravity is roughly reproduced by a naive model
based on free ${\cal N}=4$ super-Yang-Mills theory on the
world-volume.  The field theory model correctly predicts a limit on
angular momentum density, but near this limit it does not reproduce
the critical exponents one can compute from supergravity.  Analogies
with Bose condensation and modified matrix models are discussed, and a
mean field theory improvement of the naive model is suggested which
corrects the critical exponents.

\vfil
\begin{flushleft}
October 1998
\end{flushleft}
\end{titlepage}
\newpage
\section{Introduction}
\label{Introduction}

Low-energy physics on the world-volume of $N$ coincident D3-branes
\cite{jp} placed in flat ten-dimensional spacetime is dictated by
${\cal N}=4$ super-Yang-Mills theory (SYM) with gauge group $U(N)$
\cite{witBound}.  The gauge coupling and the closed string coupling
are related by $g_{YM}^2 = 2\pi g_s$.

A cluster of coincident D3-branes can also be regarded as a black
three-brane of ten-dimensional type~IIB supergravity (SUGRA)
\cite{hsBrane}.  The geometry has characteristic curvature scale $R =
(4\pi g_s N \alpha'^2)^{1/4}$.  Since the validity of supergravity
depends on having curvatures small on both the string and Planck
scales, both $N$ and $g_{YM}^2 N$ must be large.  Large 't~Hooft
coupling, $\lambda = g_{YM}^2 N \gg 1$, is the opposite limit to the
one treatable in perturbation theory, so if supergravity does capture
certain aspects of gauge theory, it is of strongly coupled gauge
theory.

One of the earliest attempts to exploit the relation between
supergravity and gauge theory as alternate low-energy descriptions of
D3-branes was the computation of the entropy \cite{gkPeet,sUnp}.  The
results disagree only by a numerical factor: in the microcanonical
ensemble (which seems the most natural from the point of view of
gravity),
  \eqn{EntMismatch}{\eqalign{
   s_{SUGRA} &= 2^{5/4} 3^{-3/4} \sqrt{\pi N} e^{3/4}  \cr
   s_{SYM} &= (4/3)^{1/4} s_{SUGRA} \ .
  }}
 In \EntMismatch, $s$ and $e$ are entropy and energy per unit
world-volume.  In supergravity, $s$ is computed using the
Bekenstein-Hawking formula, and $e$ is computed as the ADM mass
density above extremality.  In super-Yang-Mills theory, $s$ and $e$
are computed using the free theory ($g_{YM}=0$).  Conformal invariance
dictates the $e^{3/4}$ behavior but not the coefficient.  The
disagreement in the coefficient is sometimes quoted as $4/3$ rather
than $(4/3)^{1/4}$; this is the result of working in the canonical
rather than the microcanonical ensemble.  Given that the gravity and
gauge theory calculations are valid in the limits of large and small
't~Hooft coupling, respectively, the puzzle is not so much that there
is a numerical discrepancy in the coefficient but rather that the
discrepancy is finite (that is, it does not depend parametrically on
$N$ or the 't~Hooft coupling $\lambda$).  It is as if the number of
degrees of freedom in supergravity were exactly $3/4$ the number in
the free theory.  Indeed, it was pointed out in \cite{gkPeet} that the
discrepancy could be cured by excluding the gauge bosons and one
species of gauginos from the free field counting.  Because the
gauginos transform in an irreducible representation of the R-symmetry
group, this prescription breaks rotational invariance.  A more likely
scenario (recently challenged in \cite{li}) is that the coefficient on
the entropy interpolates smoothly between finite weak and strong coupling
limits.  It was shown in \cite{gkt} that at least the leading
$\alpha'$ correction to supergravity perturbed the coefficient toward
its free field theory value.

In section~\ref{Thermo} we will extend the supergravity calculation to
the case of the black three-brane with angular momentum in a plane
perpendicular to the brane.  The geometry was constructed explicitly
in \cite{Russo}, following the work of \cite{CvY}.  This spinning
three-brane has the same quantum numbers as $N$ coincident D3-branes
with a density of $R$-charge on the world-volume equal to the angular
momentum density $j$ (the $R$-symmetry group is precisely the rotation
group $SO(6)$ in the dimensions transverse to the brane).  Following
the spirit of \cite{gkPeet}, we suggest that the supergravity geometry
actually teaches us the strong-coupling thermodynamics of large-$N$
super-Yang-Mills theory.  The salient feature of that thermodynamics
is an instability for $j > N^2 T^3$ (up to numerical factors of order
unity).

It seems impossible to recover this prediction by a standard
perturbative gauge theory calculation.  Zero coupling is in a sense
the worst place to be expanding around, because in the presence of
charged massless bosons there cannot be any ``voltage'' for the charge
density $j$.  In section~\ref{NaiveField} we will suggest a formal
approach for regulating the free field partition function in the
presence of such a voltage.  We do not claim that our approach
invalidates the free field theory conclusion that massless bosons
cannot support a chemical potential; however, it may be a better
starting point from which to understand the strong coupling regime.
In particular, we shall see that near $j=0$ the regulated free energy
reproduces the supergravity results for thermodynamic quantities like
specific heat and isothermal capacitance, up to factors of order
unity.  Furthermore, it also predicts thermodynamic instability for $j
> N^2 T^3$, again in agreement with supergravity up to a finite
factor.  The critical exponents are predicted incorrectly by the
regulated field theory model, but we suggest that this might be fixed
by a mean field treatment of interactions.

Our attention will be focused, both on the supergravity side and in
the regulated field theory model, on the phase whose region of
thermodynamic stability includes the states of small $T$ and zero $j$.
We believe that the boundary of stability is actually a phase boundary
beyond which another stable phase emerges.  Our speculation is that in
this phase the D3-branes spontaneously separate.  The motivation for
this guess will be explained at the end of the paper, but the bulk of
it is devoted to three goals: 1) to demonstrate the existence of a
boundary to the region of stability of the small $j$ phase; 2) to
examine the critical properties of the thermodynamics near the
boundary; and 3) to see how the thermodynamics might be reproduced in
field theory.

\section{Spinning three-brane thermodynamics}
\label{Thermo}

Since the rotation group $SO(6)$ transverse to a three-brane in ten
dimensions has rank three, it is possible to give a three-brane three
independent (commuting) angular momenta.  We will use the solution
constructed in \cite{Russo}, which has only one angular momentum, but
there should be no difficulty in principle using the results of
\cite{CvY} to extend to the general case.  The metric is
  \eqn{RussoMet}{\eqalign{
   ds^2 &= {1 \over \sqrt{f}} \left( -h dt^2 + d\vec{x}^2 \right)  
    + \sqrt{f} \Big[ {dr^2 \over \tilde{h}} - 
     {2\ell r_0^4 \cosh\alpha \over r^4 \Delta f} \sin^2\theta 
      dt d\phi  \cr
    &\quad{} + r^2 (\Delta d\theta^2 + \tilde\Delta \sin^2\theta d\phi^2 +
      \cos^2\theta d\Omega_3^2) \Big] \ ,
  }}
 where 
  \eqn{MoreMet}{\eqalign{
   f &= 1 + {r_0^4 \sinh^2\alpha \over r^4 \Delta}  \cr
   \Delta &= 1 + {\ell^2 \cos^2\theta \over r^2}  \cr
   \tilde\Delta &= 1 + {\ell^2 \over r^2} + 
    {r_0^4 \ell^2 \sin^2\theta \over r^6 \Delta f}  \cr
   h &= 1 - {r_0^4 \over r^4 \Delta}  \cr
   \tilde{h} &= {1 \over \Delta} \left( 1 + {\ell^2 \over r^2} - 
    {r_0^4 \over r^4} \right)
  }}
 and the dilaton is constant, as for the spherically symmetric case.
Fixing the number $N$ of units of five-form flux leads to the
constraint
  \eqn{raConstraint}{
   r_0^4 \sinh\alpha \cosh\alpha = R^4 \equiv {N\kappa \over 2\pi^{5/2}}
  }
 where 
  \eqn{GravConst}{
   2\kappa^2 = 16 \pi G = (2\pi)^7 g_s^2 \alpha'^4
  }
 is the gravitational constant of type~IIB supergravity.

We define the energy of this solution as the ADM mass minus the
mass of a supersymmetric, non-rotating three-brane with the same
five-form flux: if $V$ is the spatial world-volume, then
  \eqn{EDef}{\eqalign{
   e &= {E \over V} = {M_{ADM} - M_0 \over V}  \cr 
     &= {N \sqrt{\pi} \over \kappa} \left( \tf{3}{2} \csch 2\alpha + 
      \coth 2\alpha - 1 \right)
  }}
 where we have used the BPS formula for the tension of the extremal
three-brane, $M_0/V = N\sqrt{\pi}/\kappa$.  The angular momentum
conjugate to $\phi$ can be read off from the asymptotics of
$g_{t\phi}$ \cite{mp}:
  \eqn{JDef}{
   j = {J^{xy} \over V} = {\pi^3 \over \kappa^2} \ell r_0^4 \cosh\alpha \ .
  }
 Note that $e$ has dimensions of $(\hbox{length})^{-4}$ while $j$ has
dimensions of $(\hbox{length})^{-3}$.  Finally, the entropy
  \eqn{SDef}{
   s = {A_H/4G \over V}  
     = {2\pi^4 \over \kappa^2} r_H r_0^4 \cosh\alpha
  }
 is computed from the area $A_H$ of the event horizon, whose location
$r_H$ is the unique real positive root of $\tilde{h} = 0$: $r_H^2 =
\tf{1}{2}(\sqrt{\ell^4+4r_0^4} - \ell^2)$ \cite{Russo}.

It is possible in principle to eliminate $\ell$, $r_0$, and $\alpha$
from \raConstraint, \EDef, \JDef, and \SDef, and express $s =
s(e,j;N,R)$, where we regard $(e,j)$ as variables parameterizing the
phase space and $N$ and $R = (4\pi g_s N \alpha'^2)^{1/4}$ as
parameters specifying the theory.  The nontrivial dependence on the
dimensionful parameter $R$ makes it hopeless to try to reproduce this
entropy function from ${\cal N}=4$ SYM, because ${\cal N}=4$ SYM is
conformal: it has no scale.  However, it was observed for $j=0$
already in \cite{gkPeet} that the $R$-dependence drops out in the
near-extremal limit $(M_{ADM}-M_0)/M_0 \ll 1$ (a startling consequence
is that all dependence on the 't~Hooft parameter drops out as well, in
the leading $\alpha'$ approximation to type~IIB string theory which
type~IIB supergravity represents).  For $j=0$ one is left with an
entropy function $s = s(e;N)$ which is unique up to an $N$-dependent
normalization given conformal invariance of the underlying theory and
extensivity of $S$ and $E$.  For $j\neq 0$ the entropy function again
becomes independent of $R$ in the near-extremal limit $r_0,\ell \ll
R$, but it involves a function of the dimensionless ratio $j^4/e^3$
that cannot be determined by scaling arguments.  We will now determine
it from supergravity.

The near-extremal limit has $\alpha\to\infty$, and $\alpha$ is
conveniently eliminated in favor of $r_0$ using an approximation to
\raConstraint, $e^{2\alpha} = {2N\kappa \over \pi^{5/2} r_0^4}$.
One immediately learns $e = \tf{3}{2} {\pi^3 \over \kappa^2} r_0^4$,
and with a bit more work $j = \ell \sqrt{\sqrt{\pi} N e / 3\kappa}$ 
and
  \eqn{sFinal}{
   s = s(e,j;N) = {2^{5/4} 3^{-3/4} \sqrt{\pi N} e^{3/4} \over
    \sqrt{\sqrt{1+\chi} + \sqrt{\chi}}}
     = {\sqrt{2} \pi j \over \sqrt{\sqrt{\chi(1+\chi)}+\chi}}
  }
 where 
  \eqn{chiDef}{
   \chi = {27 \pi^2 \over 8 N^2} {j^4 \over e^3} \ .
  }
 The denominators represent the nontrivial new information we are
gleaning from the spinning black hole solution.

As a safeguard against algebraic errors, and to check the
normalization of \JDef, we have computed the angular velocity $\Omega$
at the horizon.  $\Omega$ is defined by requiring the Killing vector
$\partial/\partial t + \Omega \, \partial/\partial\phi$ to be null at
the horizon.  One finds
  \eqn{CalcOmega}{
   \Omega = {\ell r_H^2 \over r_0^4 \cosh\alpha} \to
    \pi \sqrt{6} {j \over N\sqrt{e}} 
     \left( \sqrt{1+\chi} - \sqrt{\chi} \right) \ .
  }
 The first expression for $\Omega$ is exact; the second represents the
near-extremal limit.  In this limit, one can check from \sFinal\ that
  \eqn{CheckOmega}{
   \Omega = -{(\partial s/\partial j)_e \over (\partial s/\partial e)_j} \ ,
  }
 as required by the first law of black hole thermodynamics: $dE = TdS
+ \Omega dJ$.

\section{The limits of thermodynamic stability}
\label{Phase}

Our primary interest in this section is the thermodynamic stability of
near-extremal spinning D3-branes.  As a warmup, let us study the
stability of non-rotating black 3-branes which are not required to be
near-extremal.  The entropy function $s = s(e;N,R)$ has a rather
complicated form when written out explicitly.  It is simpler to refrain
from eliminating the variable $\alpha$ from \EDef\ and \SDef, since
then we have
  \eqn{etNonEx}{\eqalign{
   e &= {N^2 \over 2\pi^2 R^4} \left( \tf{3}{2} \csch 2\alpha +
    \coth 2\alpha - 1 \right)  \cr
   T &= {1 \over \pi R} {\sinh^{1/4} \alpha \over \cosh^{3/4} \alpha} \ .
  }}
 In this simple example, thermodynamic stability is equivalent to
positivity of the specific heat at constant volume, $C_V = (\partial E
/ \partial T)_V$.  In manipulations of thermodynamic quantities, we
will always hold $V$ fixed, and we will usually drop the subscript $V$
that indicates this explicitly.  Also we will work with intensive
quantities only.  Inspection of \etNonEx\ reveals that while $e$ is
strictly decreasing with $\alpha$, $T$ has a quadratic maximum at
$\alpha=\alpha_c$ where $\coth^2 \alpha_c = 3$.  This indicates a
singularity, $c_V \sim 1/\sqrt{T-T_c}$, in the specific heat, since
  \eqn{cvSing}{
   c_V = {\partial e \over \partial T} = 
    {\partial e / \partial\alpha \over \partial T / \partial\alpha} \ ,
  }
 and $T \sim T_c - (\alpha-\alpha_c)^2$ near the critical point.  For
$\alpha < \alpha_c$, $c_V < 0$, so this region is unstable.  Note that
the scale of the phase transition is set by $R$, so it can't have
anything to do with renormalizable ${\cal N}=4$ super-Yang-Mills
theory.  If it has any world-volume interpretation, it seems it would
have to be in terms of an effective, possibly non-local theory like
the Born-Infeld generalization of electromagnetism.  On the
supergravity side, the phase transition is much less arcane: it is the
standard transition from positive to negative specific heat that
occurs as mass is added to almost any charged black hole solution.
The thermodynamic instability is real in the sense that Schwarzschild
black holes really do get colder when you increase their mass.

What we will find for the near-extremal limit of spinning D3-branes is
a boundary of thermodynamic stability with the same $1/\sqrt{T-T_c}$
singularity in a specific heat, but with the critical temperature
$T_c$ set by the angular momentum density rather than a scale built
into the theory.  In other words, it is an effect which falls inside
the putative regime of validity of the renormalizable super-Yang-Mills
theory description of the world-volume.  Spinning three-branes with
$T<T_c$ are unstable, and the question we begin to address next is how
one might see this boundary of stability in field theory.

It is slightly more subtle to evaluate thermodynamic stability when
the entropy function $S = S(x_1,\ldots,x_n)$ depends non-trivially on
several extensive thermodynamic variables $x_i$.  The criterion is
sub-additivity:
  \eqn{SubAdd}{
   S(\{\lambda x_i + (1-\lambda) \tilde{x}_i\}) \geq
    \lambda S(\{x_i\}) + (1-\lambda) S(\{\tilde{x}_i\}) 
  }
 for $\lambda$ between $0$ and $1$.  When $S(\{x_i\})$ is a smooth
function, one can simply demand that the Hessian matrix $\left[
\partial^2 S / \partial x_i \partial x_j \right]$ has no positive
eigenvalues.

Applying the above considerations to the entropy function \sFinal, we
see that the condition for thermodynamic stability is $\chi \leq 1/3$.
Equivalently, we may demand that the energy function
  \eqn{MicroE}{
   e(s,j) = \tf{3}{2} {s^{4/3} \over N^{2/3} (2\pi)^{2/3}}
    \left( 1 + {4 \pi^2 j^2 \over s^2} \right)^{1/3}
  }
 derived from \sFinal\ has a positive definite Hessian.  This also
leads to $\chi \leq 1/3$, or more usefully
  \eqn{MicroSUGRAStab}{
   {j \over s} \leq {1 \over \sqrt{2} \pi} \ .
  }
 Since the Hessian matrix $\left[ \partial^2 e / \partial x_i \partial
x_j \right]$ can also be written as $\left[ \partial y_i / \partial
x_j \right]$ where $y_i = \partial e / \partial x_i$, a positive
definite Hessian means that the functions $y_i = y_i(x_1,\ldots,x_n)$
are invertible---ie that we can Legendre transform freely among the
various ensembles.  

The ensemble which will be most directly comparable to the naive field
theory model developed in section~\ref{NaiveField} is the one where
$(s,j)$ are both traded for their dual variables $(T,\Omega)$.  We
will call this the grand canonical ensemble because it is like having
indefinite particle number and a chemical potential.  The free energy
in this ensemble is usually called the grand potential, and we will
denote it as $\Xi$.  It is related to $E$ by
  \eqn{EXiLegendre}{
   E = \Xi + TS + J\Omega \ .
  }
 Written as a function $\Xi(T,\Omega,V)$ of the volume plus the
intensive quantities temperature and voltage, the grand potential is
forced by extensivity to be linear in $V$.  Indeed, $-\xi = -\Xi/V = P$
is just the pressure.  Conformal invariance implies $\xi = -\tf{1}{3}
e$.  In deriving the explicit expression
  \eqn{SUGRAXi}{\eqalign{
   \xi(T,\Omega) &= -{\pi^2 \over 8} N^2 T^4
     {\left( 1 + \pi^2 {T^2 \over \Omega^2} 
      \left( 1 - \sqrt{1 - {2 \over \pi^2} {\Omega^2 \over T^2}} 
       \right)^2 \right)^3 \over
      \left( 1 + {\pi^2 \over 2} {T^2 \over \Omega^2} 
      \left( 1 - \sqrt{1 - {2 \over \pi^2} {\Omega^2 \over T^2}} 
       \right)^2 \right)^4}  \cr
    &= -{\pi^2 \over 8} N^2 T^4 \left[ 1 + 
     {1 \over \pi^2} {\Omega^2 \over T^2} + 
     {1 \over 2\pi^4} {\Omega^4 \over T^4} + \ldots \right] \ ,
  }}
 the following ``equation of state'' is useful:
  \eqn{SUGRAeos}{
   {\Omega \over T} = {2\pi^2 j/s \over 1 + 2\pi^2 j^2/s^2} \ .
  }
 The region of allowed $\Omega/T$ is restricted: 
  \eqn{GrandSUGRAStab}{
   {\Omega \over T} \leq {\pi \over \sqrt{2}} \ ,
  }
 which as the reader can check expresses the maximum value the right
hand side of \SUGRAeos\ can attain as a function of $j/s$.  The
boundary of stability \MicroSUGRAStab\ occurs at the maximum value of
$\Omega/T$, so the \GrandSUGRAStab\ is not a good description of the
boundary of stability: the equality is obtained at the boundary, but
on both sides $\Omega/T$ decreases.\footnote{Thanks to F.~Larsen and
P.~Kraus for setting me straight on this point.}  It is possible to
quote the stability condition in the canonical ensemble, where $T$ and
$j$ are fixed: the condition is
  \eqn{jMaxSUGRA}{
   j = {9 \pi \over 16 \sqrt{2}} N^2 T^3 \ .
  }

There is really only one critical exponent in the description of the
thermodynamics near the boundary of stability.  It characterizes the
approach of $\xi$ to its value at criticality: at fixed temperature,
  \eqn{xiApproach}{
   \xi = \xi_c + j_c (\Omega_c - \Omega) - A (\Omega_c - \Omega)^{2-\gamma} + 
     O\left[ (\Omega_c - \Omega)^2 \right] \ .
  }
 where $A$ is some constant and, for the case at hand, $\gamma = 1/2$.
All other critical exponents follow from this one using scaling
properties.  In particular, the specific heat $c_\Omega$ at constant
voltage goes as $(T-T_c)^{-\gamma}$.  With $\gamma=1/2$, this is the
same $1/\sqrt{T-T_c}$ behavior we saw in the non-rotating example at
the beginning of this section.

\section{A naive field theory model}
\label{NaiveField}

Let us first recall the spectrum of ${\cal N}=4$ super-Yang-Mills
theory.  Besides the gauge bosons, which have no charge under the $R$
symmetry group $SO(6) \approx SU(4)$, there is a (color) adjoint of
Weyl fermions $\lambda^a$ in the ${\bf 4}$ of $SU(4)$ and an adjoint
of scalars $X^i$ in the ${\bf 6}$ of $SO(6)$.  Under the subgroup
$SO(2)$ of $SO(6)$ generated by $\partial/\partial\phi$, half of the
fermions (that is, $4N^2$ degrees of freedom) have charge~$1/2$ and
the other half have charge~$-1/2$, whereas of the scalars, two thirds
($4N^2$ degrees of freedom) are neutral, one sixth have charge~$1$,
and the last sixth have charge~$-1$ (note that we have included
particles and anti-particles explicitly in our counting).  All these
particles are massless since we have not Higgsed the theory.

The presence of charged massless bosons makes it impossible to apply a
``voltage'' $\Omega$ of the $SO(2)$ subgroup to the free theory,
simply because one of the thermal occupation factors $n_\pm(\vec{p}) =
1/(e^{\beta (|\vec{p}| \pm \Omega)} - 1)$ will become negative for
$|\vec{p}| < |\Omega|$.  The interactions, treated perturbatively in
the 't~Hooft coupling $\lambda$, result in a thermal mass $m \sim
\sqrt\lambda T$ to lowest order.  Thus it seems the critical value of
the potential $\Omega$ is $\Omega_c = \sqrt\lambda T$ up to a factor
of order unity: parametrically different from the supergravity result.
Factors of $\lambda$ will also enter into thermodynamic quantities
like isothermal capacitance $(\partial j/\partial\Omega)_T$ measured
at $j=\Omega=0$.  There is nothing contradictory in supposing that in
fact $\Omega_c = f(\lambda) T$ for some function $f(\lambda)$ which
approaches a constant for $\lambda \to \infty$ and $\sqrt{\lambda}$
for $\lambda \to 0$; and likewise for the various measurable
quantities like $(\partial j/\partial\Omega)_T\Big|_{\Omega=0}$.

Positing an interpolating function for every quantity that disagrees
between free field theory and supergravity may be in some sense the
correct thing to do, but it is highly non-predictive, particularly
since both the perturbation expansion in finite temperature field
theory and the $\alpha'$ expansion in string theory are difficult to
compute past the first nontrivial term.  In hopes of establishing a
better field theory understanding of the supergravity results, we will
suggest in section~\ref{Regulate} a method of regularizing the free
field theory partition function so that it becomes well-defined at
finite voltage.  In section~\ref{ThermoField} we will see that this
naive field theory treatment has some degree of success in reproducing
the phenomena observed in supergravity.  In section~\ref{MFT} we will
try to improve the naive model using mean field theory.

\subsection{A regulation scheme}
\label{Regulate}

In this section we will suggest a regulation of integrals over
$n_\pm(\vec{p})$ which makes predictions for the critical voltage
$\Omega_c$ and for $(\partial j/\partial\Omega)_T\Big|_{\Omega=0}$
which agree up to factors of order unity with the supergravity
results.  It is not our aim to revise free field theory; rather, we
hope that our procedure captures in some sense a better approximation
to the dynamics of the effective excitations of interacting
super-Yang-Mills theory.  The factor of $N^2$ on the grand potential
$\xi$ in \SUGRAXi\ is an indication that there are $N^2$ degrees of
freedom excited at finite temperature, even in the limit of strong
coupling.  We regard this as a clue that perhaps some adaptation of
the free theory can be used to approximate the statistical mechanics
of the interacting one.

The divergent integrals that occur in free field theory with a
chemical potential for massless bosons are of the form
  \eqn{LiIntOne}{
   \int_0^\infty dp \, {p^{n-1} \over e^{p - \mu} - 1}
    = \Gamma(n) \sum_{j=1}^\infty {(e^\mu)^j \over j^n} 
    = \Gamma(n) \Li_n(e^\mu)
  }
  \eqn{LiIntTwo}{
   \int_0^\infty dp \, p^{n-1} \log(1 - e^{\mu-p})
    = -\Gamma(n) \Li_{n+1}(e^\mu) \ .
  }
 where $\Li_n$ is the polylogarithm function, which is defined by the
sum in \LiIntOne\ when its argument is in the unit disk.  This
function can be analytically continued to the complex $\mu$-plane, and
it is analytic except for a branch cut along the positive real axis.
The imaginary part changes sign across this cut while the real part is
continuous.  The principle value prescription for the integrals in
\LiIntOne\ and \LiIntTwo\ for real $\mu$ is to average the values for
$\mu+i\epsilon$ and $\mu-i\epsilon$.  This amounts to replacing
$\Li_n$ by its real part $\Re\Li_n$.  One might be inclined to regard
the principle value prescription as a species of infrared regulation,
similar in spirit to differential regulation in its use of analytic
continuation to define divergent integrals.  But it differs from the
use of an explicit mass as a regulator in that it yields a
cutoff-independent free energy which is stable in a finite region
including zero voltage.  We will show that if both regularizations are
applied at once, then by making the explicit mass small one smoothly
recovers the massless case.

Using this prescription on \LiIntOne\ one can calculate the charge
density $j$ for a given potential $\Omega$ by integrating the thermal
occupation factors.  We will instead use \LiIntTwo\ to compute the
grand potential density $\xi = \Xi/V = \xi(T,\Omega;N)$, from which all
other thermodynamic quantities follow by differentiation.  For a
collection of several species of free massless charged particles in a
potential $\Omega$,
  \eqn{GenF}{
   \xi = T \int {d^3 p \over (2\pi)^3} \sum_i s_i 
    \log(1 - s_i e^{-\beta (|\vec{p}| - q_i \Omega)})
     = -{T^4 \over \pi^2} \sum_i s_i \Li_4(s_i e^{\beta q_i \Omega}) \ .
  }
 The sum over $i$ is over particle species.  For each species, $q$ is
the charge and $s = 1$ or~$-1$ according as the particle is a boson or
a fermion.  For fermions, the principle value prescription is
superfluous because the polylogarithm is evaluated at negative
argument and so is already real.  For free ${\cal N}=4$ $U(N)$ SYM,
\GenF\ specializes to
  \eqn{SYMXi}{\eqalign{
   \xi &= -{N^2 T^4 \over \pi^2} \Re\left[ 6 \Li_4(1) + 
      \Li_4(e^{-\beta\Omega}) + \Li_4(e^{\beta\Omega}) - 
      4 \Li_4(-e^{-\beta\Omega/2}) - 4 \Li_4(-e^{\beta\Omega/2}) \right]  \cr
     &= -{\pi^2 \over 8} N^2 T^4 
      \left[ {4 \over 3} + {2 \over \pi^2} {\Omega^2 \over T^2} - 
        {1 \over 4 \pi^4} {\Omega^4 \over T^4} \right]
  }}
 where we have used the identity (valid for positive integers $n$)
  \eqn{LiIdentity}{
   \Li_n(e^\mu) + (-1)^n \Li_n(e^{-\mu}) = 
    \sum_{j=0}^{\lfloor n/2 \rfloor} {2 \zeta(2j) \over
     (n-2j)!} \mu^{n-2j} \pm {i\pi \over (n-1)!} \mu^{n-1} \ .
  }
 The explicit sign on the imaginary term in \LiIdentity\ is the same
as the sign of the imaginary part of $\mu$.  From \SYMXi\ one can
extract the charge density:
  \eqn{JNFour}{
   j = -\left( {\partial \xi \over \partial\Omega} \right)_T
     = \tf{1}{2} N^2 T^3 \left[ {\Omega \over T} - 
      {1 \over 4\pi^2} {\Omega^3 \over T^3} \right] \ .
  }

As an aside, it is interesting to note that the two terms in \SYMXi\
pertaining to the charged scalars individually have singularities in
their second derivatives at the origin, but their sum is $C^\infty$.
If we were to change the charge of the scalar or give it a mass
without making the corresponding change to its anti-particle required
by CPT, this singularity would reappear.  In a non-relativistic system
of massive particles (that is, the usual Bose-Einstein setup),
  \eqn{NRBE}{\eqalign{
   j &= \int {d^3 p \over (2\pi)^3} 
       {1 \over \exp\left[ \beta \left( {p^2 \over 2m} - 
         \mu \right) \right]- 1}  \cr
     &= \left( {mT \over 2\pi} \right)^{3/2} \Re\Li_{3/2}(e^{\beta \mu})
  }}
 where we have again applied the principle value prescription to the
integral for $\mu>0$.  $\partial j/\partial\mu$ is singular at the
origin and negative for $\mu>0$, indicating thermodynamic instability.
Thus in the case of non-relativistic bosons, one recovers through the
principle value prescription the standard conclusion that the
disordered phase in a free Bose gas is stable only up to a charge
density $j = \zeta(3/2) (mT/2\pi)^{3/2}$.  

Let us now consider the case of a relativistic massive boson:
  \eqn{MRB}{
   j = \int {d^3 p \over (2\pi)^3} \left(
      {1 \over e^{\beta (\sqrt{p^2+m^2} - \Phi)} - 1} - 
      {1 \over e^{\beta (\sqrt{p^2+m^2} + \Phi)} - 1} \right) \ ,
  }
 where we have included the anti-particle in the second term.  We will
regard the mass as an explicit infrared regulator of the massless
theory.  The integral is not one of the standard special functions.
However, its singularity structure as a function of the complex
parameter $\Phi$ is easy to determine: it is analytic except for
branch cuts on the real axis for $|\Phi|>m$.  The branch cuts are
discontinuities of the imaginary part.  If one applies the principle
value prescription for real $\Phi$, one gets a continuous function
$j(\Phi)$ which is smooth except for a ``kink'' at $|\Phi|=m$, where
the function itself remains finite but its first derivative is
singular.  In standard free field theory, this kink is the signal for
Bose condensation.  But if we continue past this kink using the
principle value prescription, and if $m$ is not too large, we find a
new region where $\partial j/\partial\mu > 0$.  The size of the kink
scales as $(m/T)^{3/2}$.  In the $m \to 0$ limit, it disappears
altogether and we recover a cubic curve qualitatively identical to
\JNFour\ (it differs in exact coefficients because \JNFour\ includes
the effects of fermions).  The convergence of $j(\Phi;m)$ to
$j(\Phi;0)$ as $m \to 0$ is uniform.  If the principle value
prescription is used uniformly, physical quantities defined in a
theory with an explicit mass as the infrared cutoff have a smooth
behavior as that infrared cutoff is removed.  Note however that the
cubic curve \JNFour\ shows no sign of Bose condensation at $\Phi \to
0$.  Its thermodynamic stability will be the subject of
section~\ref{ThermoField}.

\subsection{Thermodynamics of the naive field theory model}
\label{ThermoField}

The reader who is distrustful of the regulation methods used in the
section~\ref{Regulate} may at least be willing to regard
  \eqn{SYMXiAgain}{
   \xi = -{\pi^2 \over 8} N^2 T^4 
      \left[ {4 \over 3} + {2 \over \pi^2} {\Omega^2 \over T^2} - 
        {1 \over 4 \pi^4} {\Omega^4 \over T^4} \right]
  }
 as a motivated guess for the field theory partition function.  But is
it a good guess?  In this section we will show that \SYMXiAgain\
exhibits a boundary of stability at a value of ${j \over N^2 T^3}$
differing from the supergravity result \jMaxSUGRA\ only by a finite
factor, but that the critical exponent $\gamma$ is different.  In
section~\ref{MFT} we will argue that the correct critical exponent can
be obtained through a mean field theory improvement of \SYMXiAgain.

The agreement up to factors of order unity of the first two terms in
square brackets in \SYMXiAgain\ with the corresponding terms in
\SUGRAXi\ implies a similar agreement in the small $\Omega$ limit of
all thermodynamic quantities expressible in terms of second
derivatives of the various free energies: such quantities can be
related to entries in the Hessian matrix of $\xi(T,\Omega)$, which is
  \eqn{SPHess}{
   \pmatrix{ 
    {\partial^2 \xi \over \partial T^2} & 
    {\partial^2 \xi \over \partial T \partial\Omega}  
      \cr\noalign{\vskip1\jot}
    {\partial^2 \xi \over \partial\Omega \partial T} &
    {\partial^2 \xi \over \partial\Omega^2}} = \left\{ 
     \eqalign{ & -\tf{3}{2} N^2 \pi^2 T^2 
       \pmatrix{ 1 & {\Omega \over 3 \pi^2 T} \cr\noalign{\vskip1\jot}
                 {\Omega \over 3 \pi^2 T} & {1 \over 6 \pi^2} } 
         \left( 1 + O[\Omega^2/T^2] \right) \quad
        \hbox{for SUGRA}  \cr\noalign{\vskip3\jot}
      & -\tf{3}{2} N^2 \pi^2 T^2 
       \pmatrix{ {4 \over 3} & {2 \Omega \over 3 \pi^2 T}  
        \cr\noalign{\vskip1\jot}
                 {2 \Omega \over 3 \pi^2 T} & {1 \over 3 \pi^2} }
         \left( 1 + O[\Omega^2/T^2] \right) \quad
        \hbox{for field theory.}
     } \right.
  }
 The upper left entry reflects the well-known $4/3$ disagreement of
the non-rotating case.  The $\Omega^4/T^4$ term in square brackets in
\SUGRAXi\ and \SYMXi\ are of opposite sign, and so also some of the
$O[\Omega^2/T^2]$ terms which we have not indicated explicitly in
\SPHess\ disagree in sign between supergravity and field theory.  This
is in fact the first symptom of the disagreement in critical
exponents at the boundary of stability.

It is already a nontrivial fact, however, that the regulated field
theory predicts a boundary of stability at finite $\Omega/T$.  Let us
see how it arises.  Thermodynamic stability requires that the Hessian
matrix of $\xi(T,\Omega)$ be negative definite.  This is so when
  \eqn{GrandSYMStab}{
   {\Omega \over T} \leq 2\pi \sqrt{{2 \over \sqrt{3}} - 1} \ .
  }
 Note that this is a smaller value than the one we might guess by
analogy with Bose condensation from the behavior of $j(T,\Omega)$: the
maximum of $j(T,\Omega)$ for positive $\Omega$ and fixed $T$ is
attained at $\Omega/T = 2\pi/\sqrt{3}$, which is greater than the
right hand side of \GrandSYMStab.  It is more enlightening to examine
the ``equation of state''
  \eqn{SYMeos}{
   {j \over s} = {3 \over 4\pi^2} {\Omega \over T} 
    {1 - {1 \over 4 \pi^2} {\Omega^2 \over T^2} \over
     1 + {3 \over 4 \pi^2} {\Omega^2 \over T^2}} \ .
  }
 Here the maximum of the right hand side does occur precisely at the
boundary of stability.

\begin{figure}
   \vskip-1.6cm
   \centerline{\psfig{figure=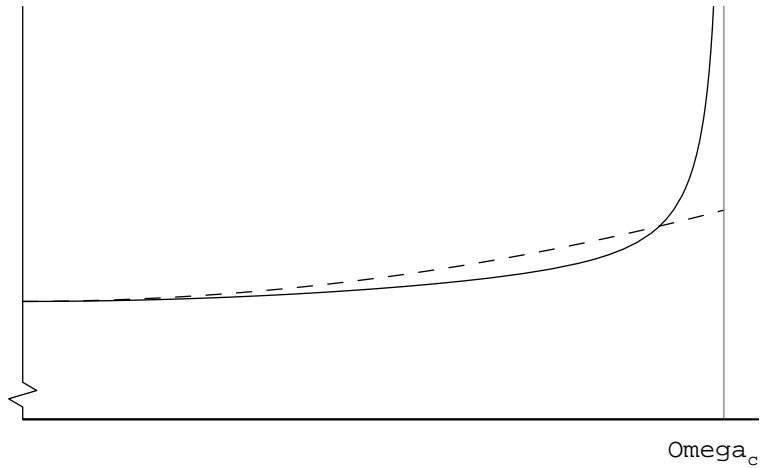}}
   \vskip-2.3cm
 \caption{$c_\Omega$ as a function of $\Omega$ for fixed $T$, in
supergravity (solid line) and in the naive field theory model (dashed
line).  The horizontal and vertical axes have been scaled differently
for the two curves so that the boundary of stability occurs at the
same point $\Omega_c$, and so that the two curves meet at $\Omega =
0$.}\label{figC}
  \end{figure}

By expanding $\xi$ in \SYMXi\ at fixed $T$ around the critical value
of $\Omega$, one finds the behavior \xiApproach\ with $\gamma=0$.  No
derivatives of $\xi$ become singular: for example, at fixed
temperature, $c_\Omega \sim (\Omega_c-\Omega)^{-\gamma} = const$.  As
one can see from figure~\ref{figC}, $c_\Omega$ smoothly approaches a
maximum as $\Omega \to \Omega_c$.

\subsection{Interactions and modified critical behavior}
\label{MFT}

The comparison between supergravity and the naive field theory model
is a little like the comparison between superfluidity in ${}^4{\rm
He}$ and Bose condensation in an ideal gas \cite{London,Wilks}.  The
free field treatment of Bose condensation predicts the location of the
${}^4{\rm He}$ phase transition to within $30\%$.  But whereas in the
free field treatment the specific heat has a finite peak at the
transition temperature, in real ${}^4{\rm He}$ it exhibits a
$\log(T-T_c)$ divergence.  The interactions modify the critical
behavior.

There is also an example of logarithmic corrections to critical
scaling in an exactly soluble model of two-dimensional quantum
gravity: the modified matrix models
\cite{ddsw,Sugino,gkMod,klebMod,khMod}.  In matrix models one
typically has an expansion
  \eqn{MMFree}{
   {\cal F} \sim \hbox{(analytic in $\Delta$)} + \Delta^{2-\gamma} + 
    \hbox{(higher order in $\Delta$)} \ .
  }
 for the free energy.  Here $\Delta$ is the cosmological constant,
which is taken to zero to obtain large random surfaces, and $\gamma$
is termed the string susceptibility.  The standard $c=1$ model has
$\Delta^2/\log\Delta$ as the leading nonanalytic term.  One can
introduce a $(\tr\Phi^2)^2$ term, which can be regarded as an
interaction among otherwise free fermions in a potential well.  The
positions of the fermions are the eigenvalues of the matrix $\Phi$.
If the strength of the interaction is tuned appropriately, it alters
the critical behavior of $c=1$ to $\Delta^2 \log\Delta$ (and,
incidentally, changes the location of the critical point in the
coupling space of the matrix model by a finite factor).  This is like
going from $\gamma$ barely negative to $\gamma$ barely positive.  For
$c<1$, the shift is finite, and again changes the sign of $\gamma$:
upon adding the interaction, $\gamma \to \bar\gamma =
\gamma/(\gamma-1)$.

It is our hope that by including an effective description of the
interactions in our field theory model, we may see a similar
modification in the critical behavior---optimistically, one which will
predict $\gamma = 1/2$.

Let us try to evaluate whether this is a realistic hope by considering
a mean field theory treatment.  One can suppose that the presence of a
density $j$ of $R$-charge sets up a ``self-field'' contribution to the
voltage.\footnote{This assumption may seem strange in view of the fact
that $SO(6)$ $R$-symmetry is not gauged in the world-volume theory.
It is however gauged in the supergravity.  If the dynamics of
supergravity are thought to be contained in the strongly coupled gauge
theory, then it seems quite natural for two $R$-charges to feel each
other's influence.}  In order to work in terms of dimensionless
quantities (as conformal invariance demands), we assume a relation
  \eqn{OmegaTotal}{
   {\Omega_{\rm tot} \over T} = {\Omega \over T} + 
     f\left( {j \over N^2 T^3} \right)
  }
 where $f(\rho)$ is some function to be determined.  We obtain a
consistency condition by replacing $j$ in \OmegaTotal\ with the
integral over thermal occupation factors:
  \eqn{JReplace}{
   j \to \int {d^3 p \over (2\pi)^3} \sum_i q_i 
    {1 \over e^{\beta (|\vec{p}| - q_i \Omega_{\rm tot})} - s_i}
  }
 where as before $i$ indexes particles, $q_i$ is the charge, and $s_i
= 1$ for bosons and $-1$ for fermions.  The integral is regulated
using the principle value prescription.  \JReplace\ boils down to
replacing $\Omega$ by $\Omega_{\rm tot}$ in \JNFour\ and then plugging the
resulting expression for $j$ back into \OmegaTotal.  Note however that
the physical charge density must be obtained by differentiating the
grand potential $\xi$ by the applied voltage $\Omega$.  The grand
potential is obtained by replacing $\Omega$ by $\Omega_{\rm tot}$ in \SYMXi.
Using the consistency condition one can write $\xi$ wholly in terms of
$\Omega$ and $T$.

To shorten the formulas, let us introduce some notation for
dimensionless quantities:
  \def\omegat{\omega_{\rm tot}}
  \def\omegatc{\omega_{{\rm tot},c}}
  \eqn{DimLessQuant}{
   {j \over N^2 T^3} = \rho \qquad
   {\Omega \over T} = \omega \qquad
   {\Omega_{\rm tot} \over T} = \omegat \ .
  }
 Then the grand potential can be written as
  \eqn{Xirewrite}{
   -{\xi \over N^2 T^4} \equiv \tilde{g}(\omega) = g(\omegat) \equiv
    {\pi^2 \over 6} + \tf{1}{4} \omegat^2 - {1 \over 32 \pi^2} \omegat^4 \ ,
  }
 and the consistency condition is
  \eqn{Consistent}{
   \omega = \omegat - f(g'(\omegat)) = 
    \omegat - f\left( \tf{1}{2} \omegat - 
      {1 \over 8 \pi^2} \omegat^3 \right) \ .
  }
 Primes will always denote the differentiation of a function with
respect to its argument.

The quantity $\mu = \partial\omegat/\partial\omega$ can be regarded in
analogy with $\partial B/\partial H$ in magnetic systems as a
permeability: $\mu > 1$ is the analog of paramagnetism, where the
self-field contribution reinforces the applied voltage, and $\mu < 1$
is the analog of diamagnetism.  Let $\mu_0$ be the permeability at
zero voltage.  Modulo a constraint on the behavior of $f(\rho)$ for
finite $\rho$, we claim that $\gamma = 1/2$ arises precisely when
$\mu_0 = 2/3$, and otherwise $\gamma = 0$ or $3/2$ according as
$\mu_0 > 2/3$ or $\mu_0 < 2/3$.  The constraint is that $f(\rho)$ must
be such that
  \eqn{SuperConvex}{
   4 \tilde{g}(\omega) \tilde{g}''(\omega) - 3 \tilde{g}'(\omega)^2 > 0
  }
 for $\omega$ less than the value where $\omegat = \omegatc \equiv 2\pi$.
This is a special point because $g'(2\pi) = 0$.  We will refer to
\SuperConvex\ as the convexity constraint because it is precisely the
requirement that the Hessian matrix of $\xi(T,\Omega)$ be negative
definite.

First note by explicit differentiation of \Consistent\ that $\mu_0 =
2/3$ is equivalent to $f'(0) = -1$.  Let us assume $\mu_0 = 2/3$.
Then $\partial\omega/\partial\omegat = 0$ at $\omegat = \omegatc$.
Near $\omegat = \omegatc$ we have
  \eqn{OmegaCrit}{\eqalign{
   \omega_c - \omega &= a_2 (\omegatc-\omegat)^2 + a_3 (\omegatc-\omegat)^3 + 
     O[(\omegatc-\omegat)^4]  \cr
   g(\omegat) &= g_0 + g_2 (\omegatc-\omegat)^2 + g_3 (\omegatc-\omegat)^3 + 
     O[(\omegatc-\omegat)^4] \ .
  }}
 Plugging the first equation into the second, one obtains
  \eqn{GTCrit}{
   \tilde{g}(\omega) = \tilde{g}_0 + \tilde{g}_1 (\omega_c-\omega) + 
    \tilde{g}_{3/2} (\omega_c-\omega)^{3/2} + O[(\omega_c-\omega)^2]
  }
 which indicates that indeed $\gamma=1/2$.  The reader interested in
the details will have no trouble verifying 
  \eqn{Allai}{\seqalign{\span\TL & \span\TR\quad & \span\TL & \span\TR\quad &
    \span\TL & \span\TR}{
  \omega_c &= \omegatc = 2\pi &&&&  \cr
  a_2 &= {3 \over 4\pi} &
   a_3 &= \tf{1}{6} f'''(0) - {1 \over 8\pi^2} &&  \cr
  g_0 &= {2 \pi^2 \over 3} &
   g_2 &= -\tf{1}{2} &
   g_3 &= {1 \over 4\pi}  \cr
  \tilde{g}_0 &= {2 \pi^2 \over 3} &
   \tilde{g}_1 &= -{2\pi \over 3} &
   \tilde{g}_{3/2} &= 
     {4 \over 27} \sqrt{\pi \over 3} (3 + 2 \pi^2 f'''(0)) \ .
  }}
 In \Allai\ we have used the fact that $f(-\rho) = -f(\rho)$ by
symmetry, so all even derivatives of $f(\rho)$ vanish at the origin.
The critical charge density is given by 
  \eqn{jMaxMFT}{
   {j \over N^2 T^3} = \rho_c = -\tilde{g}_1 = {2 \pi \over 3} \ .
  }

For $\mu_0 < 2/3$, $\partial\omega/\partial\omegat < 0$ at $\omegat =
\omegatc$.  This makes it impossible to invert \Consistent\ to obtain
$\omegat$ as a single-valued function of $\omega$ past the maximum
value for $\omega$, which is obtained at some value $\omegat =
\omega_{{\rm tot},m} < \omegatc$.  Because $g'(\omega_{{\rm tot},m})
\neq 0$, this scenario leads to $\gamma = 3/2$.  For $\mu_0 > 2/3$,
$\omegatc$ could still be the critical point if \SuperConvex\ is not
violated earlier, but there would be a linear term in
$\omegatc-\omegat$ in the expression for $\omega_c-\omega$ in
\OmegaCrit, so $\tilde{g}(\omega)$ would be analytic in $\omega$ near
$\omega_c$: $\gamma = 0$, as we saw with no interactions included at
all.  So only at the special value $\mu_0 = 2/3$ can we see the
$\gamma = 1/2$ behavior predicted by supergravity.  The precise shape
of $\tilde{g}(\omega)$ depends on the behavior of $f(\rho)$ for finite
$\rho$.  But the location $\omega_c = 2\pi$ of the boundary of
stability and the prediction $\gamma=1/2$ rely only on $f'(0) =
-1$---again, modulo the convexity requirement \SuperConvex.  It may be
entertaining for the reader to note that with $\mu_0 = 2/3$ the
discrepancy between the isothermal capacitance in field theory and
supergravity is a factor of $8/9$, which is numerically an improvement
over the factor of $2$ that we saw for $\mu_0=1$ in
\SPHess.

At this point we must ask if it is indeed possible to satisfy
\SuperConvex\ for $0 < \omegat < 2\pi$, because if \SuperConvex\ is
violated for $\omegat < \omegatc = 2\pi$, then the system becomes
thermodynamically unstable ``by accident'' before one can come to the
special point $\omegat = \omegatc$, and all our analysis will have
been for nothing.  As long as $\omega$ is a strictly increasing
function of $\omegat$, \SuperConvex\ is equivalent to requiring that
  \eqn{SCrewrite}{
   {\partial \over \partial \omegat} \left[ {1 \over g(\omegat)} \left( 
     {\partial g/\partial\omegat \over \partial\omega/\partial\omegat} 
    \right) \right] > 0 \ ,
  }
 so a necessary condition (and also a sufficient one if
$\partial\omega/\partial\omegat > 0$ can be shown) is to require
\SCrewrite\ for $0 < \omegat < 2\pi$.  We do not have complete results
on what restrictions this implies on the behavior of $f(\rho)$.
However, it is straightforward to verify that linear functions
$f(\rho) = 2 {\mu_0-1 \over \mu_0} \rho$ satisfy the requirement if
and only if $\mu_0 = 2/3$.  For $\mu_0 < 2/3$, linear $f(\rho)$
results in $\gamma=3/2$ in the way described in the previous
paragraph.  For $\mu_0 > 2/3$, in fact the critical point occurs for
$\omegat < \omegatc = 2\pi$, and $\gamma=0$ by an analysis
qualitatively identical to the case with no interactions at all.

It is conceivable that through some peculiar singular behavior of
$f(\rho)$ at finite $\rho$ one could arrange $\gamma = 1/2$ and a
smaller critical value for $\omegat$.  Of course, such a setup would
seem extremely ad hoc unless one had a justification for the
particular form of $f(\rho)$ from an analysis of the interactions.
Such an analysis is highly desirable at this point, at least for small
$\rho$.  It would be particularly amusing if the strong coupling limit
$\lambda \to \infty$ managed somehow to tune $\mu_0$ to $2/3$.
Indirect evidence for whether this is happening might be extracted
from a study of $\alpha'$ corrections in the supergravity, along the
lines of \cite{gkt}.

\section{Conclusions}
\label{Conclusion}

We have investigated the thermodynamics of coincident D3-branes
spinning in a plane orthogonal to the branes.  Using the standard
techniques of black hole thermodynamics we obtained an expression for
the entropy which indicates thermodynamic stability only up to a
special value of the dimensionless ratio of the angular momentum
density and the temperature.  We can write all thermodynamic functions
in terms of intensive quantities because conformal invariance forbids
the presence of any dimensionful parameters: any function of
dimensionful variables can be re-expressed in terms of intensive
quantities by scaling each variable by the appropriate power of the
volume.  The absence of a scale means that physical relations like an
equation of state or the location of a boundary of stability must be
expressible in terms of dimensionless ratios of intensive quantities.
It also implies that only one critical exponent, $\gamma$, describes
the approach of the world-volume theory to the boundary of stability.
It for instance gives the scaling law $c_\Omega \sim
(T-T_c)^{-\gamma}$ at fixed voltage (as explained above, voltage
$\Omega$ is the thermodynamic variable dual to $R$-charge).  We have
found $\gamma = 1/2$ from supergravity.

In the superconformal field theory which characterizes low-energy
fluctuations of the branes, angular momentum is interpreted as
electric charge under a subgroup of the $R$-symmetry group.  We have
suggested an alteration of free field theory in which one can
calculate very simply the critical value of the angular momentum
density where the phase transition takes place.  The calculation
employs the thermodynamic ensemble where voltage rather than
$R$-charge density is held fixed.  This ensemble is formally
ill-defined when there are free massless charged bosons.  The
alteration of free field theory which we have suggested relies on a
regulation procedure employing analytic continuation to render all the
relevant integrals finite.  The reader is entitled to question the
validity of our regulation scheme: negative thermal occupation number
is a serious pathology.  Nevertheless, if supergravity is supposed to
be teaching us about super-Yang-Mills theory at strong coupling, then
spinning black holes show that the theory is stable up to voltages on
the order of the temperature.  A description of the interacting theory
has to start somewhere, and our starting point, the regulated free
field theory, enjoys a measure of success.  It leads to an expression
for the grand potential which exhibits a boundary of stability similar
to the one observed in supergravity, and at a similar value of the
charge density at fixed temperature.

The regulated field theory wrongly predicts $\gamma = 0$.  In analogy
to Bose condensation and the modified matrix models, we have suggested
that including an effective description of the interactions may change
the scaling behavior.  We have substantiated this claim by showing in
mean field theory that if the three-branes respond ``diamagnetically''
to an applied voltage (that is, if the total voltage on the branes is
reduced from the applied voltage by a self-field contribution) then
the scaling behavior can be modified to $\gamma=1/2$.  One obtains
$\gamma=1/2$ at a special value, $\mu_0 = 2/3$, of the
``permeability.''  This value separates a region of $\gamma=0$
behavior ($\mu_0 > 2/3$) from a region of $\gamma=3/2$ behavior
($\mu_0 < 2/3$).  The regulated free field theory and the mean field
treatment of interactions are not motivated by any deep understanding
of strong coupling gauge dynamics, but rather by a desire to see in a
simple way how and whether finite-temperature field theory can
reproduce the thermodynamics of spinning near-extremal D3-branes.

The analysis of this paper leaves us at a point similar to the state
of matrix models of two-dimensional quantum gravity: we have a
description of critical behavior near a boundary, but on the other
side of the boundary our description breaks down.  In the matrix
models, the intuition was that eigenvalues were leaking out of the
local potential well and falling down a $V(\lambda) \sim -\lambda^4$
hill.  A clear description of the physics seemed difficult because it
demanded a ``non-perturbative'' definition of two-dimensional quantum
gravity.  In Bose condensation a new stable phase (the Bose
condensate) emerges on the other side of the boundary.  The obvious
analog of Bose condensation on D3-branes is Higgsing: some of the
scalars develop expectation values.  This suggests that the cluster of
coincident branes should be splitting up, perhaps into a ring
(condensed state) around a central core (normal state).  A
supergravity solution of this type may be difficult to write down
explicitly.  If it could be shown that there is a correlation length
in the scalar fields on the uncondensed side of the phase transition
which diverges at the boundary, we would regard it as evidence in
favor of the Higgsing scenario.  We hope to report on these issues in
the future.

\section*{Note Added}

After this work was completed, we received the paper \cite{cort} which
has a tangential relation to section~\ref{Thermo} of this paper.  In
\cite{cort} the Euclidean version of the spinning three-brane solution
was considered.  The thermodynamics is qualitatively different because
the angular momentum and velocity are imaginary.  The entropy
function that follows from (3.24) of \cite{cort} can be obtained
formally from our results by replacing $\sqrt{\chi} \to -\sqrt{\chi}$
in the first expression for $s$ in \sFinal.

\section*{Acknowledgements}

It is a pleasure to thank A.~Strominger for suggesting this line of
research, for pointing out to me reference \cite{Russo}, and for many
useful discussions.  I profited from comments from D.~Fisher,
B.~Halperin, D.~Nelson, and C.~Thorn, and I thank I.~Klebanov for
reading an early draft of the paper.  This research was supported by
the Harvard Society of Fellows, and also in part by the National
Science Foundation under grant number PHY-98-02709.

\newpage

\bibliography{spin}
\bibliographystyle{ssg}

\end{document}